\documentclass[12pt, preprint,preprintnumbers,nofootinbib, groupedaddress,superscriptaddress,amsmath,amssymb]{revtex4}
\usepackage{graphicx}
\usepackage{dcolumn}
\usepackage{bm}
\usepackage{amssymb}
\usepackage{amsmath}
\usepackage{epsfig}    
\usepackage{color}
\usepackage{hhline}

\def\be{\begin{equation}}
\def\ee{\end{equation}}
\newcommand{\bea}{\begin{eqnarray}}
\newcommand{\eea}{\end{eqnarray}}
\newcommand{\nn}{\nonumber}

\numberwithin{equation}{section}

\begin{document}
{\begin{flushright}{KIAS-P19033,  APCTP Pre2019 - 014}\end{flushright}}

\title{A two loop induced neutrino mass model \\ with modular $A_4$ symmetry}

\author{Takaaki Nomura}
\email{nomura@kias.re.kr}
\affiliation{School of Physics, KIAS, Seoul 02455, Korea}

\author{Hiroshi Okada}
\email{hiroshi.okada@apctp.org}
\affiliation{Asia Pacific Center for Theoretical Physics (APCTP) - Headquarters San 31, Hyoja-dong,
Nam-gu, Pohang 790-784, Korea}

\date{\today}

\begin{abstract}
We propose a model with radiatively induced neutrino mass at two-loop level, applying modular $A_4$ symmetry.
The neutrino mass matrix is formulated where the structure of associated couplings are restricted by the symmetry.
Then we show several predictions in the lepton sector, satisfying lepton flavor violations as well as neutrino oscillation data.
We also discuss muon anomalous magnetic moment and briefly comment on dark matter candidate. 
\end{abstract}
\maketitle
\newpage

\section{Introduction}
Understanding flavor structure for quark and lepton sectors is still important even after the discovery of Higgs boson in the standard model (SM). Along the idea of thought, non-Abelian discrete groups have been vastly adopted as flavor symmetries in order to predict and/or reproduce current experimental data of the mixings and masses of quarks and leptons~\cite{Altarelli:2010gt,
Ishimori:2010au,Ishimori:2012zz,Hernandez:2012ra,King:2013eh,King:2014nza, King:2017guk,Petcov:2017ggy}.
However, since there are too many possibilities of their applications, one cannot address something concrete model independently in this manner.
Recently, modular originated flavor symmetries have been proposed by~\cite{Feruglio:2017spp, deAdelhartToorop:2011re}
that are more promising ideas to obtain predictions to the quark and lepton sector, since Yukawa couplings also have a representation of the flavor groups.  
Their typical groups are found in basis of  the $A_4$ modular group \cite{Feruglio:2017spp, Criado:2018thu, Kobayashi:2018scp, Okada:2018yrn, Nomura:2019jxj, Okada:2019uoy, deAnda:2018ecu, Novichkov:2018yse}, $S_3$ \cite{Kobayashi:2018vbk, Kobayashi:2018wkl}, $S_4$ \cite{Penedo:2018nmg, Novichkov:2018ovf}, $A_5$ \cite{Novichkov:2018nkm, Ding:2019xna}, larger groups~\cite{Baur:2019kwi}, multiple modular symmetries~\cite{deMedeirosVarzielas:2019cyj}, double covering groups of several modular symmetries~\cite{Liu:2019khw, Liu:2020akv, Yao:2020zml, Kikuchi:2020nxn} that have been applied to studies of  flavor structures of quarks and leptons and the feature of dark matter (DM) candidate~\cite{Nomura:2019jxj}.
Also, an interesting application to the generalized CP transformations has been appeared in, e.g., refs.~\cite{Baur:2019iai, Kobayashi:2019uyt, Novichkov:2019sqv, Holthausen:2012dk, Yao:2020qyy}.
Another advantage of this modular groups is that fields and couplings have to assign a modular weight originated from modular groups. This number can be identified to be a symmetry to stabilize DM candidate if DM is included in a model.

In this paper, we apply an $A_4$ modular symmetry to the lepton sector in a framework of modified Zee-Babu type model~\cite{Zee-Babu, Okada:2014qsa} generating non-zero masses of neutrinos at two-loop level. In the model, we introduce exotic vector-like charged leptons in addition to the field contents in original Zee-Babu model~\cite{Zee-Babu} which propagate inside a loop diagram generating neutrino mass. In our analysis, we show several predictions to the lepton sector, satisfying constraints of lepton flavor violations (LFVs) as well as neutrino oscillation data and discussing muon anomalous magnetic moment (muon $g-2$). Finally, we briefly comment on our DM candidate in the conclusion.~\footnote{A DM candidate has been discussed in non-Abelian discrete symmetries in refs.~\cite{Hirsch:2010ru, Lamprea:2016egz, delaVega:2018cnx}, in which a symmetry to stabilize DM is originated from a remnant symmetry after breaking of flavor symmetry.}

This paper is organized as follows.
In Sec.~\ref{sec:realization},   we give our model set up under the modular $A_4$ symmetry, writing down relevant fields  and couplings and their assignments. 
Then, we formulate the valid Lagrangians, Higgs potential, exotic field mass matrix, LFVs, muon $g-2$, neutrinos mass matrix, 
and numerical analysis in which we show several predictions to satisfy all the data that we will discuss.
Finally we conclude and discuss in Sec.~\ref{sec:conclusion}.

\begin{center} 
\begin{table}[tb]
\begin{tabular}{|c||c|c|c||c|c|c|c|c||}\hline\hline  
&\multicolumn{3}{c||}{ Fermions} & \multicolumn{5}{c||}{Scalars} \\\hline
  & ~$L_{L_{e,\mu,\tau}}$~& ~$\ell_{R_{e,\mu,\tau}}$~ & ~$E_{i}$~ & ~$H$~& ~$\eta$~ & ~$S^-$~ & ~$k^{++}$~& ~$\varphi$~
  \\\hline 
 $SU(2)_L$ & $\bm{2}$  & $\bm{1}$  & $\bm{1}$ & $\bm{2}$   & $\bm{2}$  & $\bm{1}$ & $\bm{1}$& $\bm{1}$  \\\hline 
$U(1)_Y$ & $-\frac12$ & $-1$ & $-1$  & $\frac12$& $\frac12$ & $-1$  & $2$ & $0$      \\\hline
 $A_4$ & $1,1',1''$ & $1,1',1''$ & $3$ & $1$ & $1$ & $1$ & $1$  & $1$      \\\hline
 $-k_I$ & $0$ & $0$ & $-1$ & $0$ & $-1$ & $-1$ & $-2$& $-2$   \\\hline
\end{tabular}
\caption{Field contents of fermions and scalar fields
and their charge assignments under $SU(2)_L\times U(1)_Y\times A_{4}$ in the lepton and boson sector, where $k$ is the number of modular weight, $a=1,2,3$, and the quark sector is the same as the SM.}
\label{tab:fields}
\end{table}
\end{center}

\section{ Model} 
\label{sec:realization}
Here we explain our model with modular $A_4$ symmetry in which some fields have non-zero modular weight and couplings with nonzero modular weight are modular forms.
In the fermion sector, we introduce three exotic singly-charged leptons $E$ as a triplet under $A_4$ with modular weight $-1$,
while all the SM leptons $L_{L}\equiv[e_L,\nu_L]^T,\ell_{R}$ have zero modular weight and are assigned three kinds of singlet ${1,1',1''}$ for each flavor under $A_4$. 
In the scalar sector, we introduce an isospin doublet field $\eta$ and three singlet fields $(\varphi, S^-,k^{++})$ having non-zero modular weights $-1$, $(-2,-1,-2)$ {and hypercharges $1/2$, $(0,-1,2)$} respectively,
where all the scalar fields are true singlets under $A_4$. 
We assume $\eta$ to be an inert boson
\footnote{In practice, $\eta$ is not needed to be inert, since only the singly-charged component contributes to our neutrino mass at two-loop level. If $\eta$ has nonzero VEV, we have mixing between the SM charged-leptons $\ell$ and exotic ones $E$ that is proportional to the ratio defined as $\ell$ mass divided by that of $E$. The mixing should be small enough to satisfy the experimental data since it can affect SM lepton gauge interactions and would induce LFV process~\cite{Crivellin:2020ebi,Patrignani:2016xqp}. Therefore, VEV of $\eta$ has to be suppressed, and thus we do not consider this kind of case in our paper.},
and its neutral component can be a DM candidate whose stability is assured by nonzero modular weight; {this is due to the fact that all couplings should have even modular weight and fields with odd modular weight cannot singly appear in interactions. } 
Vacuum expectation value (VEV) of $H$ and $\varphi$ is respectively denoted by $v_H/\sqrt2$ and $v_\varphi/\sqrt2$, where $H$ is identified as SM-like Higgs field.
We summarize field assignments in table \ref{tab:fields}. 
Under these symmetries, one writes renormalizable Lagrangian as follows:
\begin{align}
-&{\cal L}_{Lepton} =
\sum_{\ell=e,\mu,\tau} y_{\ell}\bar L_{L_\ell} e_{R_{\ell}} H\nn\\
&+\kappa_{1} \bar L_{L_1} (f_{1} E_{R_1} + f_{2} E_{R_3}+ f_{3}  E_{R_2} )\eta
\nn\\
&+\kappa_{2}  \bar L_{L_2} (f_{3} E_{R_3} + f_{1} E_{R_2}+ f_{2} E_{R_1} )\eta
\nn\\
&+\kappa_{3}  \bar L_{L_3} (f_{2} E_{R_2} + f_{1}\bar E_{R_3}+ f_{3} E_{R_1} )\eta
\nn\\
&+ \frac{\rho_1^{L,R}}3
\left[ f'_{1} (2\overline{ E_{1}^C} E_1 -\overline{ E_2^C} E_3-\overline{ E_3^C} E_2)_{LL,RR}
+f'_{2} (2\overline{ E_2^C} E_2 -\overline{ E_1^C} E_3-\overline{ E_3^C} E_1)_{LL,RR}\right.\nn\\
&\left.
\hspace{1.5cm}+f'_{3} (2\overline{ E_3^C} E_3 -\overline{E_1^C} E_2-\overline{E_2^C} E_1)_{LL,RR}\right]
 k^{++}\nn\\
& +\rho_2^{L,R}  Y^{(4)}_{\rm1}  (\overline{E_1^C} E_1 +\overline{E_2^C} E_3 + \overline{E_3^C} E_2)_{LL,RR} k^{++}
 +\rho_3^{L,R}  Y^{(4)}_{\rm1'}  (\overline{ E_2^C} E_2 +\overline{E_1^C} E_3 + \overline{E_3^C} E_1)_{LL,RR} k^{++} \nn\\
&+ M_{0} (\bar E_{1} E_{1} + \bar E_{2} E_{2}+ \bar E_{3} E_{3})_{LR}\\
&
+ \frac{\alpha_E}3 \left[f_1(2\bar E_1E_1-\bar E_2E_2-\bar E_3E_3)+f_2(2\bar E_3E_2-\bar E_1E_3-\bar E_2E_1)
+f_3(2\bar E_2E_3-\bar E_1E_2-\bar E_3E_1)\right]_{LR}\varphi \nn\\
&
+ \frac{\beta_E}2 \left[f_1(\bar E_3E_3-\bar E_2E_2)+f_2(2\bar E_2E_1-\bar E_1E_3)
+f_3(\bar E_1E_2- \bar E_3E_1)\right]_{LR}\varphi
+  {\rm h.c.}, \label{eq:lag-lep}
\end{align}
where $Y^{(2)}_3\equiv(f_1,f_2,f_3)^T$ is $A_4$ triplet with modular weight $2$, $Y^{(4)}_3\equiv(f'_1,f'_2,f'_3)^T$ is $A_4$ triplet with modular weight $4$, $Y^{(4)}_{1,1'}$ are $A_4$ singlets with modular weight $4$, the parameters $M_0$, $\alpha_E$ and $\beta_E$~\footnote{ We would like thank referee to point these terms out.} include $1/(-i\tau+i\bar\tau)$ factor implicitly to make the term modular invariant, 
and the charged-lepton mass eigenstate is directly given by the first term.
Thus, the observed mixing matrix for lepton sector is found in the neutrino sector only. 
The  modular forms of weight 2, {$(f_{1},f_{2},f_{3})$},  transforming
as a triplet of $A_4$, is written in terms of Dedekind eta-function $\xi(\tau)$ and its derivative $\xi'(\tau)$ \cite{Feruglio:2017spp}:
\begin{eqnarray} 
\label{eq:Y-A4}
f_{1}(\tau) &=& \frac{i}{2\pi}\left( \frac{\xi'(\tau/3)}{\xi(\tau/3)}  +\frac{\xi'((\tau +1)/3)}{\xi((\tau+1)/3)}  
+\frac{\xi'((\tau +2)/3)}{\xi((\tau+2)/3)} - \frac{27\xi'(3\tau)}{\xi(3\tau)}  \right), \nonumber \\
f_{2}(\tau) &=& \frac{-i}{\pi}\left( \frac{\xi'(\tau/3)}{\xi(\tau/3)}  +\omega^2\frac{\xi'((\tau +1)/3)}{\xi((\tau+1)/3)}  
+\omega \frac{\xi'((\tau +2)/3)}{\xi((\tau+2)/3)}  \right) , \label{eq:Yi} \\ 
f_{3}(\tau) &=& \frac{-i}{\pi}\left( \frac{\xi'(\tau/3)}{\xi(\tau/3)}  +\omega\frac{\xi'((\tau +1)/3)}{\xi((\tau+1)/3)}  
+\omega^2 \frac{\xi'((\tau +2)/3)}{\xi((\tau+2)/3)}  \right)\,.
\nonumber
\end{eqnarray}
%
The overall coefficient in Eq. (\ref{eq:Yi}) is 
one possible choice; it cannot be uniquely determined. Thus we just impose the purtabative limit {$f_{1,2,3}\lesssim\sqrt{4\pi}$} in the numerical analysis.
Once $Y^{(2)}_3$ is found, the other Yukawa couplings are constructed by multiplication rules of $A_4$ as follows:
$Y^{(4)}_3\equiv(f_1^2-f_2 f_3,f_3^2-f_1f_2,f_2^2-f_1f_3)^T$, $Y^{(4)}_1=f_1^2+2 f_2f_3$, and  $Y^{(4)}_{1'}=f_3^2+2 f_1f_2$, where $Y^{(4)}_{1''}$ vanishes due to the relation $f_2^2+2f_1 f_3=0$.

$A_4$ modular invariant Higgs potential is given by
\begin{align}
{\cal V} &= -\mu_H^2 |H|^2+\mu_\eta^2 |\eta|^2-\mu_\varphi^2 |\varphi|^2+ \mu_k^2 |k^{++}|^2 + \mu_S^2 |S^{-}|^2 
+ \mu_{ssk} Y^{(4)}_1 S^-S^-k^{++}  +\lambda Y^{(4)}_1 (H^Ti \sigma_2 \eta) S^- \varphi\nn\\
&+\lambda_H|H|^4  
+ \lambda_\eta |\eta|^4 
+\lambda_\varphi |\varphi|^4 + \lambda_k |k^{++}|^4 + \lambda_S |S^-|^4 
+  \lambda_{H\eta} |H|^2|\eta|^2+  \lambda'_{H\eta} |H^\dag\eta|^2 \nn\\
&+  \lambda_{H\varphi} |H|^2|\varphi|^2
+  \lambda_{Hk} |H|^2|k^{++}|^2
+  \lambda_{HS} |H|^2|S^-|^2
+  \lambda_{\eta \varphi} |\eta|^2|\varphi|^2+  \lambda_{\eta k} |\eta|^2 |k^{++}|^2+  \lambda_{\eta s} |\eta|^2|S^+|^2\nn\\
&
+ \lambda_{\varphi k} |\varphi|^2 |k^{++}|^2
+ \lambda_{\varphi s} |\varphi|^2 |S^{-}|^2
+ \lambda_{ks} |k^{++}|^2 |S^-|^2 
+ {\rm h.c.}
, \label{eq:pot}
\end{align}
where $\sigma_2$ is the second component of  Pauli matrices, all the couplings except ($\mu_{ssk},\ \lambda,\ \mu_H,\lambda_H$) include the modular invariant factor $1/(-i\tau+ i \bar\tau)^{k_{I}} (k_I=1,2)$ depending on modular weight of the corresponding term.
Here, we define each of the scalar components as follows:
\begin{align}
H&=
\left[\begin{array}{c}
h^+  \\ 
\frac{v_H+ h+i z}{\sqrt2}  \\ 
\end{array}\right], \
\eta=
\left[\begin{array}{c}
\eta^+  \\ 
\frac{\eta_R+i \eta_I}{\sqrt2}  \\ 
\end{array}\right], \
\varphi = \frac{v_\varphi+ \varphi_R+i \varphi_I}{\sqrt2},
\end{align}
where $v_H=246$ GeV, $h^+$ and $z$ are respectively absorbed by the $W$ and $Z$ boson in the SM. Solving stable conditions; $\partial V/\partial h=0,\ \partial V/\partial \varphi_R =0 $, we write VEVs in terms of coefficients of Higgs potential as follows: $v_H= -\frac{-\lambda_{H\varphi}\varphi^2+2 \mu_H^2}{\sqrt{2}\lambda_H}$ and $v_\varphi= -\frac{-\lambda_{H\varphi}\varphi^2+2 \mu_\varphi^2}{\sqrt{2}\lambda_\varphi}$. To evade global minimum for charged bosons of $S^+,k^{++}$, we have to impose the following condition:
\begin{align}
|\mu_{ssk}|<2\sqrt{\Lambda}  \mu,
\end{align}
where $\Lambda\equiv \lambda + \sum_{i=H,\eta,\cdots}\lambda_i$ is sum of all the quartic couplings, and
$\mu\equiv \sqrt{-\mu_H^2-\mu_\varphi^2+\mu_\eta^2+\cdots}$ is sum of all the quadratic terms. Here, we assume all the scalar fields to be the same direction~\cite{Nishiwaki:2015iqa}.
Furthermore, we show the inert condition for $\eta$. To assure the potential be bounded below, following conditions are required:
 \begin{align}
&-\frac{2}{\sqrt{21}}\sqrt{\lambda_H\lambda_\eta} < \lambda_{H\eta}+\lambda'_{H\eta},\quad 
-\sqrt6 \lambda'_{H\eta}<|\lambda_{H\eta}+\lambda'_{H\eta}|,\\
& -\frac{2}{\sqrt{21}}\sqrt{\lambda_\eta\lambda_\varphi} < \lambda_{\eta\varphi},\quad
 -\frac{2}{\sqrt{21}}\sqrt{\lambda_\eta\lambda_S} < \lambda_{\eta S}, \quad
 -\frac{1}{\sqrt{21}}\sqrt{\lambda_\eta\lambda_k} < \lambda_{\eta k}, \\
&-\frac12  \sqrt{\lambda_{\eta\varphi} \lambda_{HS}} <\lambda< \frac12  \sqrt{\lambda_{\eta\varphi} \lambda_{HS}},\quad
-\frac1{\sqrt3}  \sqrt{\lambda_{H\varphi} \lambda_{\eta S}} <\lambda< \frac1{\sqrt3}   \sqrt{\lambda_{H\varphi} \lambda_{\eta S}},
\end{align}
where $0<\lambda_{H,\eta,\varphi,S,k}$. Under these conditions, the minimum of $V$ is stable and global  when all the mass squared eigenstates are positive.
In the singly-charged bosons in basis of  $S^\pm,\eta^\pm$, which mix each other through $\lambda$, we define the mixing and its mass eigenvalue as follows:
\begin{align}
& S^{\pm}=c_\alpha H_1^{\pm} + s_\alpha H_2^{\pm},\quad \eta^{\pm} =-s_\alpha H_1^{\pm} + c_\alpha H_2^{\pm},
\quad s_{2\alpha}=\frac{2 {\lambda} v_H v_\varphi }{m_{H_2}^2 - m_{H_1}^2},\\
&
m_{H_1}^2 = \frac12\left[m_{SS}^2+m_{\eta\eta}^2-\sqrt{m_{SS}^4+m_{\eta\eta}^4 -2 m_{SS}^2m_{\eta\eta}^2+4m_{\eta S}^4}\right],
\nn\\
&
m_{H_2}^2 = \frac12\left[m_{SS}^2+m_{\eta\eta}^2+\sqrt{m_{SS}^4+m_{\eta\eta}^4 -2 m_{SS}^2m_{\eta\eta}^2+4m_{\eta S}^4}\right],
\end{align}
where $s_\alpha(c_\alpha)$ is the short-hand symbol of $\sin \alpha(\cos \alpha)$, and 
\begin{align}
& m_{SS}^2 = \frac{\lambda_{H S}}{2} v_H^2+ \frac{\lambda_{\varphi S}}{2} v_\varphi^2 +\mu_S^2,\\
& m_{\eta\eta}^2 = \frac{\lambda_{H \eta}}{2} v_H^2+ \frac{\lambda_{\eta \varphi}}{2} v_\varphi^2 +\mu_\eta^2,\\
& m_{\eta S}^2 =- \frac{\lambda_{}}{2} v_H v_\varphi,
 \end{align}

After the electroweak spontaneous symmetry breaking,  the charged-lepton mass matrix is given by
\begin{align}
m_\ell&= \frac {v_H}{\sqrt{2}}
\left[\begin{array}{ccc}
y_e  & 0 & 0 \\ 
0 & y_\mu &  0 \\ 
0 & 0  & y_\tau  \\ 
\end{array}\right]\equiv
\left[\begin{array}{ccc}
m_e  & 0 & 0 \\ 
0 & m_\mu &  0 \\ 
0 & 0  & m_\tau  \\ 
\end{array}\right].
\end{align}

 The mass matrix of  $E$ is 
given by 
\begin{align}
{\cal M}_E&=
M_0\left[\begin{array}{ccc}
1 &0 &0 \\ 
0& 1 & 0  \\ 
0& 0 &  1    \\ 
\end{array}\right]
+
\frac{\alpha_E v_\varphi}{3\sqrt2}
\left[\begin{array}{ccc}
2f_1 &-f_3 &-f_2 \\ 
-f_2& -f_1 & 2 f_3  \\ 
-f_3& 2 f_2 &  f_1    \\ 
\end{array}\right]
+
\frac{\beta_Ev_\varphi}{2\sqrt2}
\left[\begin{array}{ccc}
0 &f_3 &-f_2 \\ 
f_2& -f_1 & 0  \\ 
-f_3& 0 &  f_1    \\ 
\end{array}\right]\nn\\
&\approx
M_0\left[\begin{array}{ccc}
1 &0 &0 \\ 
0& 1 & 0  \\ 
0& 0 &  1    \\ 
\end{array}\right]
+
\frac{\alpha_E v_\varphi}{3\sqrt2}
\left[\begin{array}{ccc}
2 &0 &0 \\ 
0& -1 & 0  \\ 
0& 0 &  1    \\ 
\end{array}\right]
+
\frac{\beta_E v_\varphi}{2\sqrt2}
\left[\begin{array}{ccc}
0 &0 &0 \\ 
0& -1 & 0  \\ 
0& 0 &  1    \\ 
\end{array}\right]
,\label{eq:mn}
\end{align}
where we have used the following approximations that there are hierarchies among $f_1\sim1 >> f_2>>f_3$ at the fundamental region of $\tau$.
Moreover, $\alpha_E,\beta_E$ includes $1/(-i\tau+\bar\tau)$ whose absolute value is 0.5 at most.
One can evaluate that the second and third mass terms are subdominant by $\sim20\%$ compared to $M_0$, when $M_0=v_\varphi$.
Thus, we assume in our numerical analysis that ${\cal M}_E$ is diagonal with degenerate mass eigenstates $M_0$.
%

{\it LFVs, Muon g-2, Neutrino masses}; 
First of all, let us rewrite the term of $f$ as
\begin{align}
-{\cal L}_f&=\bar L_{L_i} F_{ij} E_{R_j} \eta+\bar \nu_{L_i} F_{ij} E_{R_j}(-s_\alpha H_1^{+} + c_\alpha H_2^{+}) +{\rm h.c.},\\
F& \equiv
\left[\begin{array}{ccc} \kappa_1& 0 & 0 \\ 0 & \kappa_2 & 0   \\ 0 & 0 & \kappa_3 \\  \end{array}\right]
\left[\begin{array}{ccc} f_1 & f_3 & f_2 \\ f_2 & f_1 & f_3   \\ f_3 & f_2 & f_1   \\  \end{array}\right].
\end{align}
This interaction induces LFV decay process $\ell_i \to \ell_j \gamma$ at one-loop level, and
the branching ratios are given by 
\begin{align}
&BR(\ell_i\to \ell_j\gamma)\approx \frac{48\pi^3\alpha_{em} G_{ij}}{G^2_F}\left(1+\frac{m^2_j}{m^2_i}\right) |{\cal A}_{ij}|^2,\\
&{\cal A}_{ij}=-\frac{F_{ja}F^\dag_{ai}}{(4\pi)^2} G(m_{\eta_0},M_a),\\
&G(m_a,m_b)\approx\frac{m^6_a-6m^4_am^2_b+3m^2_am^4_b+ 2 m^6_b+12m^2_a m^4_b\ln\left(\frac{m_b}{m_a}\right)}{12(m^2_a-m^2_b)^4},
\end{align}
{where $G_{21}=1$, $G_{31}=0.1784$, $G_{32}=0.1736$, $\alpha_{em}$ is the electromagnetic fine structure constant, and $G_F$ is the Fermi constant. }
In addition, the muon $g-2$ is given by
\begin{align}
\Delta a_\mu\approx -2m_\mu {\cal A}_{22}.
\end{align}
The experimental upper bounds for LFVs are given by~\cite{TheMEG:2016wtm, Aubert:2009ag,Renga:2018fpd}
\begin{align}
{\rm BR}(\mu\to e\gamma)\lesssim 4.2\times10^{-13},\quad 
{\rm BR}(\tau\to e\gamma)\lesssim 3.3\times10^{-8},\quad
{\rm BR}(\tau\to\mu\gamma)\lesssim 4.4\times10^{-8},\label{eq:lfvs-cond}
\end{align}
which will be imposed in our numerical calculation.
In addition the experimental result of muon $g-2$ is found as $\Delta a_\mu=(26.1\pm8)\times10^{-10}$ with positive value
at 3.3 $\sigma$ deviation from the SM prediction.

\begin{figure}[tb]\begin{center}
\includegraphics[width=77mm]{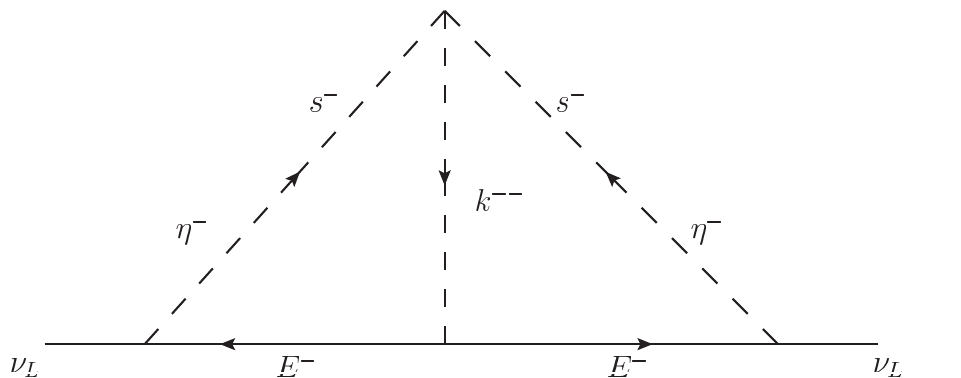}
\caption{Neutrino mass diagram at two-loop level.}   
\label{fig:neut}\end{center}\end{figure}

{\it Neutrino mass matrix} is given at two-loop level by the diagram in Fig.\ref{fig:neut}, and its formula is found as
\begin{align}
& (m_\nu)_{ij} \approx \sum_{a,b} F_{ia} {\cal M}_{ab} F^T_{bj} +  (\rm transpose) ,\\
&{\cal M}_{ab} \equiv  \mu_{ssk} s^2_\alpha c^2_\alpha \frac{ g^{L*}_{ab} +g^{R*}_{ab} }{(4\pi)^4}
\sum_{I,J}^{1,2}\int_0^1{[dx]_3\int_0^1[dx']_3} \left[\frac{2}{z-1}\ln\left[\frac{\Delta_{ab}^{IJ}}{z^2-z}\right] - \frac{M_a M_b}{\Delta_{ab}^{IJ}}\right] ,\\
&\Delta_{ab}^{IJ}\equiv -x' (x M_a^2 +y m_I^2 + z m_k^2) + (z^2-z)(y' M_b^2+z' m_J^2),
\end{align}
where $m_k$ is the mass eigenstate of $k^{\pm\pm}$, $M_{a,b}$ is mass eigenstate of $E$ that is supposed to be degenerate and denoted by $M_0$, $[dx]_3\equiv dxdydz\delta(1-x-y-z)$,
and here we assume to be $g\equiv g^{L}=g^{R}$ for simplicity. Then $g$ is given by
\begin{align}
g&=
\left[\begin{array}{ccc}
\frac23\rho_1 f'_1+\rho_2Y^{(4)}_1 &-\frac13\rho_1 f'_3 & -\frac13\rho_1 f'_2 +\rho_3 Y^{(4)}_{1'} \\ 
-\frac13\rho_1 f'_3& \frac23\rho_1 f'_2+\rho_3Y^{(4)}_{1'} &  -\frac13\rho_1 f'_1 +\rho_2 Y^{(4)}_{1} \\ 
 -\frac13\rho_1 f'_2 +\rho_3 Y^{(4)}_{1'} &   -\frac13\rho_1 f'_1 +\rho_2 Y^{(4)}_{1}  &   \frac23\rho_1 f'_3   \\ 
\end{array}\right]\label{eq:mn},
\end{align}
where $\rho_{1,2,3}\equiv \rho^L_{1,2,3}=\rho^R_{1,2,3}$.
Then $m_\nu$ is diagonalized by $U_\nu m_\nu U_\nu^T\equiv {\rm diag}[m_1,m_2,m_3]$, where $\sum_{i=1,2,3} m_{i}\lesssim 0.12$ eV is given by the recent cosmological data~\cite{Aghanim:2018eyx}.
Since the charged-lepton is mass eigenstate from the beginning, one identifies $U_\nu$ as $U_{MNS}$.
Once we define ${\cal M}\equiv \mu_{ssk}\tilde {\cal M}$, atmospheric mass difference squared $\Delta m^2_{atm}$ is written as
\begin{align}
\Delta m^2_{atm} = \mu^2_{ssk}(\tilde m_3^2-\tilde m_1^2)\quad {\rm NH},\\
\Delta m^2_{atm} = \mu^2_{ssk}(\tilde m_2^2-\tilde m_3^2)\quad {\rm IH},
\end{align}
where $\tilde m_{1,2,3}\equiv m_{1,2,3}/\mu_{ssk}$.
It suggests that $\Delta m_{atm}$  is always fitted to the experimental value by controlling $\mu_{ssk}$.
Thus, we consider   $\Delta m_{atm}$ as input parameter that corresponds to experimental value.
Then, the solar mass difference square is given by
\begin{align}
&\Delta m_{\rm sol}^2=\Delta m^2_{atm} \frac{{\tilde m_{2}^2-\tilde m_{1}^2}}{\tilde m_3^2-\tilde m_1^2}
\quad {\rm NH},\\
&\Delta m_{\rm sol}^2=\Delta m^2_{atm} \frac{{\tilde m_{2}^2-\tilde m_{1}^2}}{\tilde m_2^2-\tilde m_3^2}
\quad {\rm IH}
,
 \end{align}
 where both the above equations can uniformly be written by $\Delta m_{\rm sol}^2=\mu_{ssk}^2 ({\tilde m_{2}^2-\tilde m_{1}^2})$.
Each of mixing is given in terms of the component of $U_{MNS}$ as follows:
\begin{align}
\sin^2\theta_{13}=|(U_{MNS})_{13}|^2,\quad 
\sin^2\theta_{23}=\frac{|(U_{MNS})_{23}|^2}{1-|(U_{MNS})_{13}|^2},\quad 
\sin^2\theta_{12}=\frac{|(U_{MNS})_{12}|^2}{1-|(U_{MNS})_{13}|^2}.
\end{align}
Also, the effective mass for the neutrinoless double beta decay is given by
\begin{align}
m_{ee}=\mu_{ssk}|\tilde m_1 \cos^2\theta_{12} \cos^2\theta_{13}+\tilde m_2 \sin^2\theta_{12} \cos^2\theta_{13}e^{i\alpha_{21}}
+\tilde m_3\sin^2\theta_{13}e^{i(\alpha_{31}-2\delta_{CP})}|,
\end{align}
where its observed value could be measured by KamLAND-Zen in future~\cite{KamLAND-Zen:2016pfg}. 
The effective mass for $\beta$ decay $\langle m_\beta\rangle(\equiv \sum_{a=1,2,3}m_a |(U_{MNS})_{1a}|^2)$is given by~\cite{Avignone:2007fu, Farzan:2002zq, Huang:2019tdh} 
\begin{align}
\langle m_\beta\rangle=|\mu_{ssk}|\sqrt{(\tilde m_1^2 \cos^2\theta_{12} \cos^2\theta_{13}+\tilde m_2^2 \sin^2\theta_{12} \cos^2\theta_{13} +\tilde m_3^2 \sin^2\theta_{13})},
\end{align}
where its observed value could be measured by future experiments such as KATRIN~\cite{Huang:2019tdh}. 
{\it Notice here that we can consider all the Yukawa couplings as real parameters except two of $\rho_{1,2,3}$ due to phase redefinition of fields. Here, we take $\rho_1$ to be real.}

\section{Numerical analysis}
In this section, we show numerical analysis to satisfy all of the constraints that we discussed above,
 where we consider both normal and inverted hierarchy of neutrino mass ordering.
First of all, we provide the allowed ranges for neutrino mixings and mass difference squares at 3$\sigma$ range~\cite{Esteban:2018azc} for NH(IH) as follows:
\begin{align}
&\Delta m^2_{\rm atm}=[2.431-2.622 (2.413-2.606)]\times 10^{-3}\ {\rm eV}^2,\
\Delta m^2_{\rm sol}=[6.79-8.01]\times 10^{-5}\ {\rm eV}^2, \nonumber \\
&\sin^2\theta_{13}=[0.02044-0.02437 (0.02067-0.02461)],\ 
\sin^2\theta_{23}=[0.428-0.624 (0.433-0.623)], \nonumber \\  
& \sin^2\theta_{12}=[0.275-0.350],\label{eq:exp}
\end{align}
where the values without bracket indicate the same range for both NH and IH.
The free dimensionless parameters $\kappa_i, \rho_i$ (i=1-3)
are taken to be the range of $[0.1-1]$,
while the mass parameters $M_0, m_{k}, m_\eta, m_{H_{1,2}}$ are $[1-5]$ TeV, where $m_\eta$ $m_k$ are respectively the masses of $\eta_0$ and $k^{\pm\pm}$.

\begin{figure}[tb]\begin{center}
\includegraphics[width=75mm]{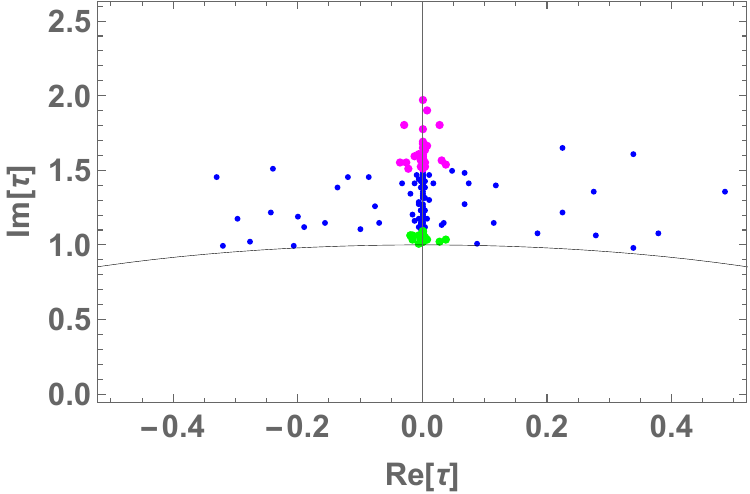}
\includegraphics[width=75mm]{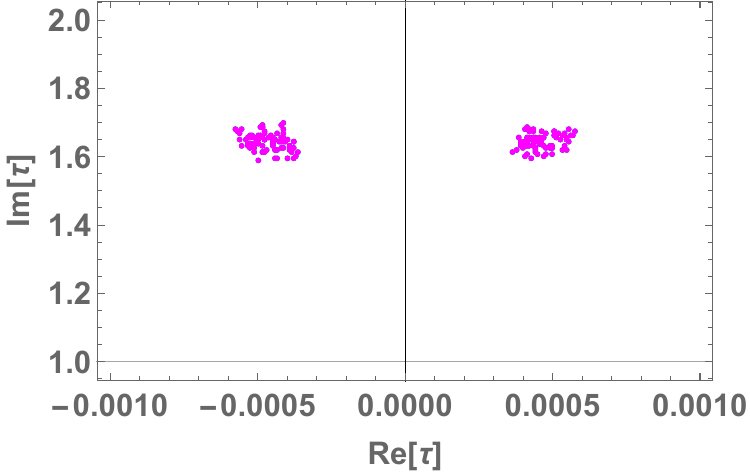}
\caption{Allowed region in terms of the real part of $\tau$ and the imaginary one of $\tau$, where the left figure is the one for  NH and the right one IH.
The blue color represents whole the region of $\tau$ in the fundamental region,
the green one is nearby $\tau=i$, and the magenta one is nearby $\tau=i\times\infty$.} 
\label{fig:tau}\end{center}\end{figure}
Fig.~\ref{fig:tau} shows  allowed region in terms of  the real and the imaginary part of $\tau$, where the left figure is the one for  NH and the right one is for IH.
The blue color represents whole the region of $\tau$ in the fundamental region, the green one is nearby $\tau=i$, and the magenta one is nearby $\tau=i\times\infty$.
We could not find any allowed region nearby $\tau=\omega$ for both the cases. The IH case has only the region nearby $\tau=i\times\infty$.
 %

\begin{figure}[tb]\begin{center}
\includegraphics[width=75mm]{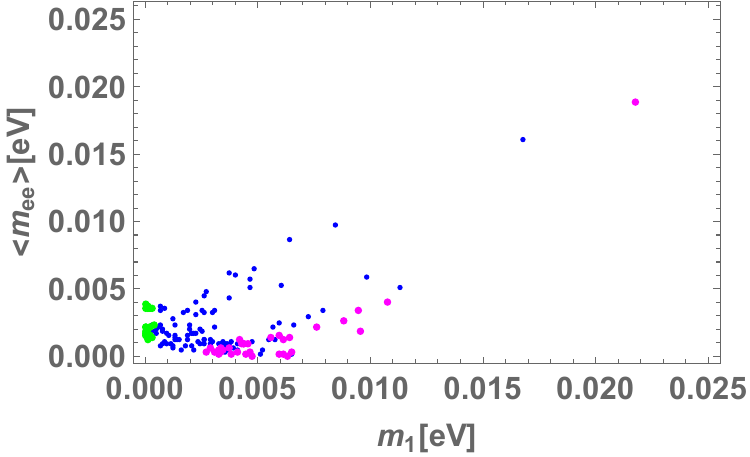}
\includegraphics[width=75mm]{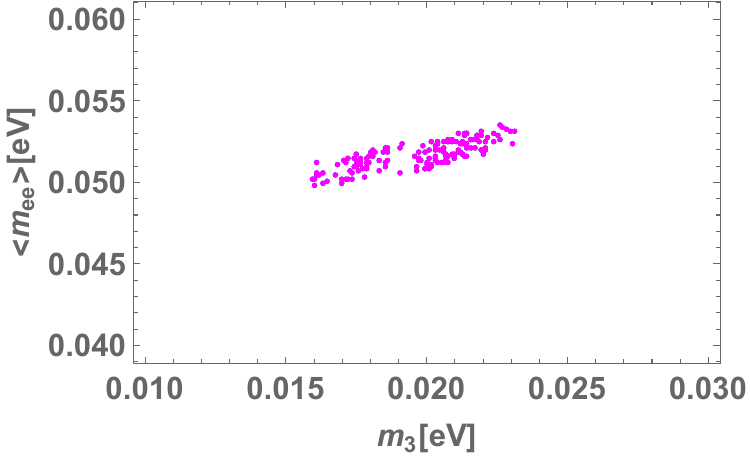}
\includegraphics[width=75mm]{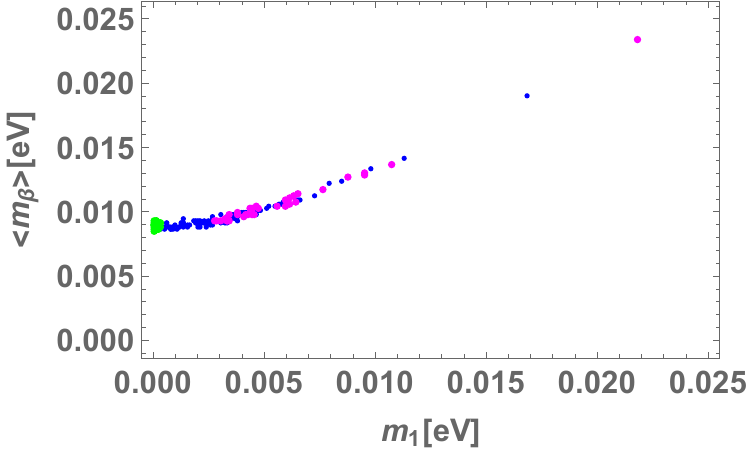}
\includegraphics[width=75mm]{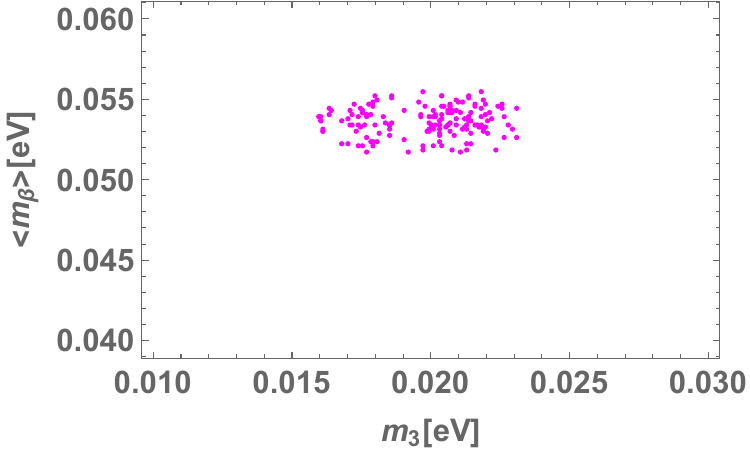}
\caption{Allowed region in terms of the lightest active neutrino mass $m_1(m_3)$ and the neutrinoless double beta decay $\langle m_{ee}\rangle$(the effective mass for $\beta$ decay $\langle m_\beta\rangle$) for up(down) side, where the left figure is the one for  NH and the right one IH, and the figure configuration and its colors are the same as those of Fig.\ref{fig:tau}.} 
\label{fig:sins}\end{center}\end{figure}
Fig.~\ref{fig:sins} shows  allowed region in terms of the lightest active neutrino mass $m_1(m_3)$ and the neutrinoless double beta decay $\langle m_{ee}\rangle$(the effective mass for $\beta$ decay $\langle m_\beta\rangle$) for up(down) side, where the left figure is the one for  NH and the right one IH, and the figure configuration and its colors are the same as those of Fig.\ref{fig:tau}.
We could not find any allowed region nearby $\tau=\omega$ for both the cases. The IH case has only the region nearby $\tau=i\times\infty$.
These figures suggest that $m_{1(3)}=[0-0.022]([0.016-0.035])$ eV for NH(IH), and $\langle m_{ee}\rangle=[0-0.019]([0.050-0.053])$ eV and $\langle m_{\beta}\rangle=[0-0.018]([0.052-0.056])$ eV for NH(IH).
Furthermore, we predict $m_1\sim0$ eV and $\langle m_{ee}\rangle=[0.001-0.004]$ eV nearby $\tau=i$, while
 $m_1=[0.0025-0.022]$ eV and $\langle m_{ee}\rangle=[0-0.019]$ eV nearby $\tau=i\times\infty$.

\begin{figure}[tb]\begin{center}
\includegraphics[width=75mm]{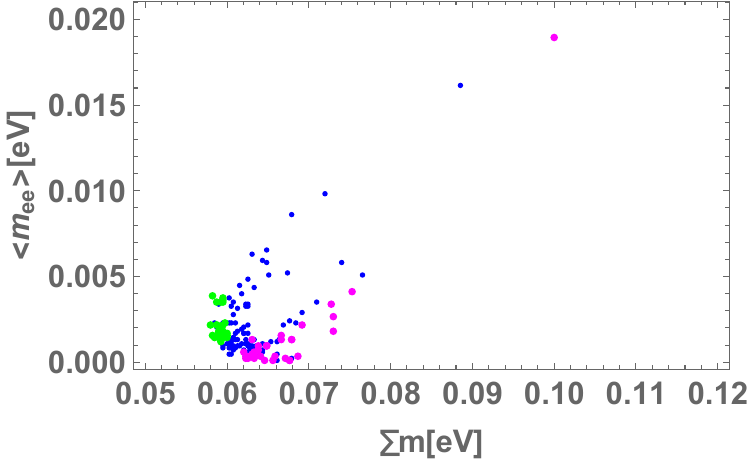}
\includegraphics[width=75mm]{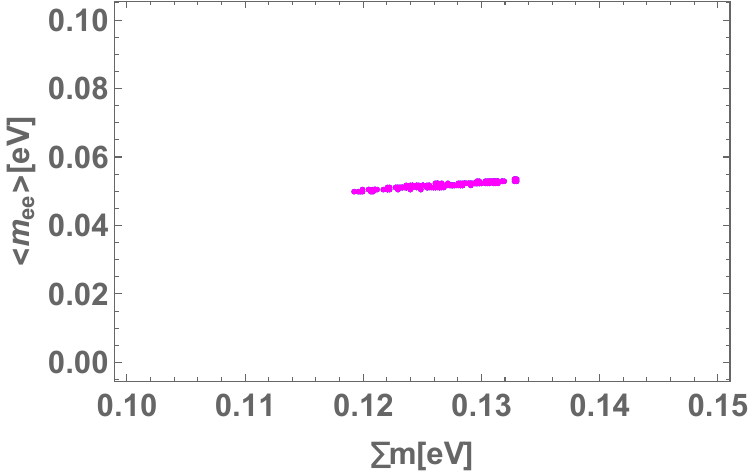}
\includegraphics[width=75mm]{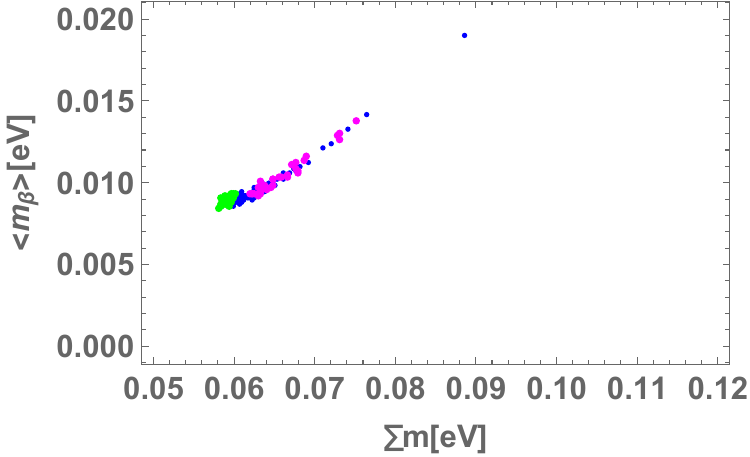}
\includegraphics[width=75mm]{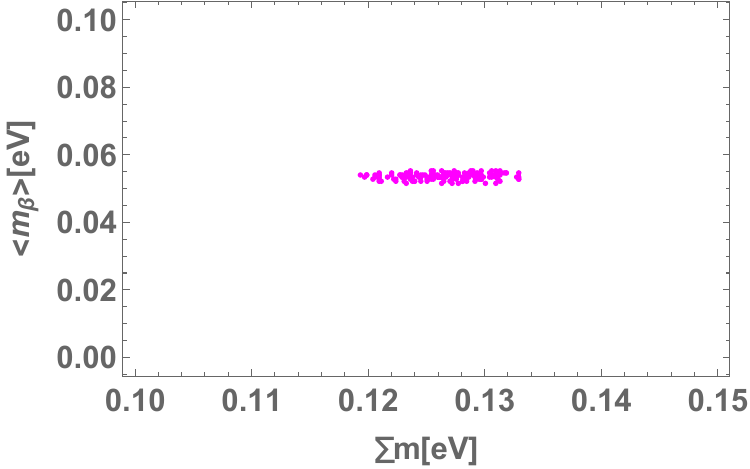}
\caption{Allowed region in terms of the sum of neutrino masses($\equiv \sum m$ eV) and the neutrinoless double beta decay $\langle m_{ee}\rangle$(the effective mass for $\beta$ decay $\langle m_\beta\rangle$) for up(down) side, where the figure configuration and its colors are the same as those of Fig.\ref{fig:tau}.}  
\label{fig:lep1}\end{center}\end{figure}
Fig.~\ref{fig:lep1} shows allowed region in terms of the sum of neutrino masses and the neutrinoless double beta decay $\langle m_{ee}\rangle$(the effective mass for $\beta$ decay $\langle m_\beta\rangle$) for up(down) side, where the figure configuration and its colors are the same as those of Fig.\ref{fig:tau}.
These figures suggest that $\sum m=[0.058-0.1]([0.119-0.134])$ eV for NH(IH) for NH(IH).
Furthermore, we predict $\sum m=[0.058-0.061]$ eV nearby $\tau=i$, while
 $\sum m=[0.062-0.1]$ eV nearby $\tau=i\times\infty$.

\begin{figure}[tb]\begin{center}
\includegraphics[width=75mm]{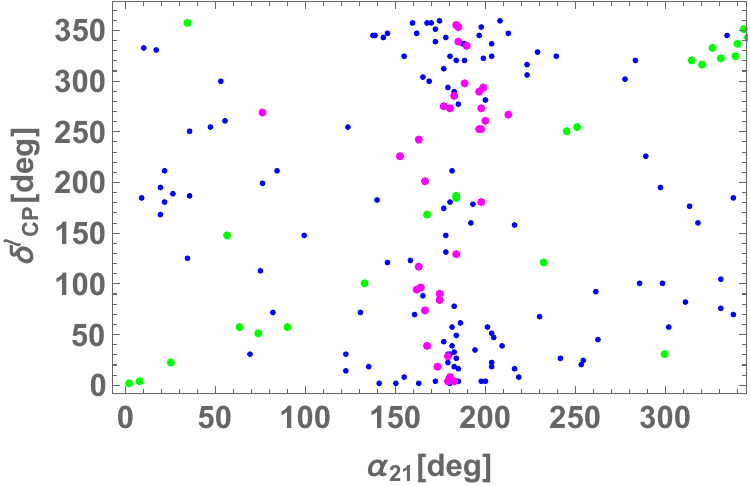}
\includegraphics[width=75mm]{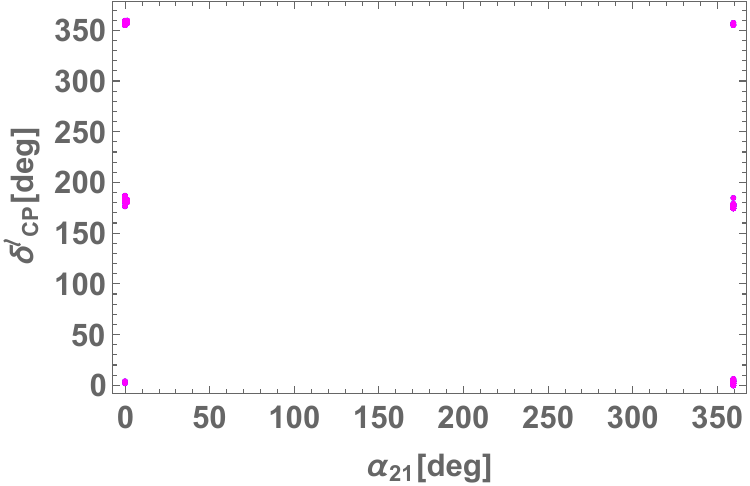}
\caption{Allowed region in terms of a Majorana phase of $\alpha_{21}$ and Dirac CP phase $\delta_{CP}^\ell$, where the figure configuration and its colors are the same as those of Fig.\ref{fig:sins}.}   
\label{fig:phase1}\end{center}\end{figure}
Fig.~\ref{fig:phase1} shows allowed region in terms of a Majorana phase of $\alpha_{21}$ and Dirac CP phase $\delta_{CP}^\ell$, where the figure configuration and its colors are the same as those of Fig.\ref{fig:tau}.
In case of NH, these figures suggest any value of Dirac CP phase is allowed, being independent of $\tau$.
The $\alpha_{21}$ tends to be localized at $180^{\circ}$ nearby $\tau=i\times \infty$, while any value of $\alpha_{21}$ is allowed for the other region of $\tau$.  
In case of IH, only the values $0^{\circ}$ and $180^{\circ}$ are allowed for $\alpha_{21}$ and $\delta^\ell_{CP}$.

\begin{figure}[tb]\begin{center}
\includegraphics[width=75mm]{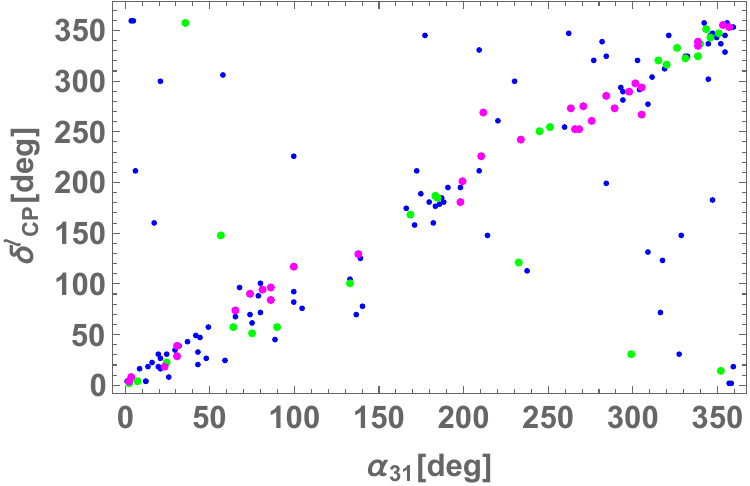}
\includegraphics[width=75mm]{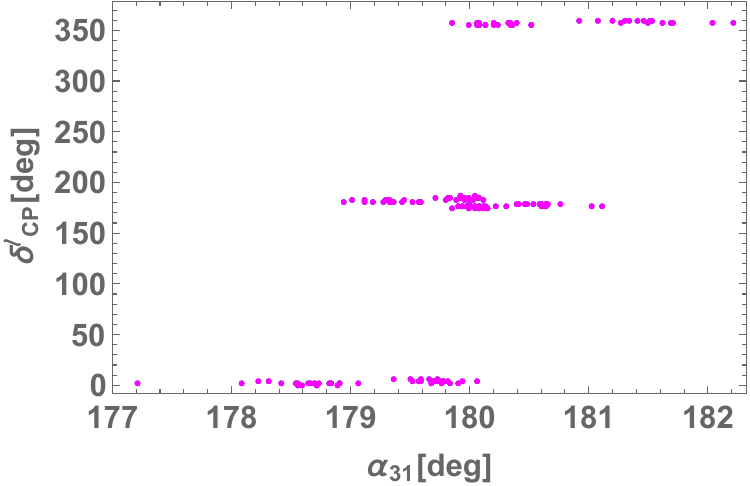}
\caption{Allowed region in terms of a Majorana phase of $\alpha_{31}$ and Dirac CP phase $\delta_{CP}^\ell$, where the figure configuration and its colors are the same as those of Fig.\ref{fig:tau}.}   
\label{fig:phase2}\end{center}\end{figure}
Fig.~\ref{fig:phase2} shows allowed region in terms of a Majorana phase of $\alpha_{31}$ and Dirac CP phase $\delta_{CP}^\ell$, where the figure configuration and its colors are the same as those of Fig.\ref{fig:sins}.
In case of NH,
These figures suggest $\alpha_{31}$ tends to be linearly proportional to $\delta^\ell_{CP}$, although all the region is allowed.
In case of IH,  although all the region for $\alpha_{31}$ is allowed, $\delta^\ell_{CP}$ seems to be fixed.
In Table~\ref{samplelepton}, we show two successful samples  for NH and one sample for IH, where two samples of NH corresponds to the region at nearby $\tau=i$ and $\tau=i\infty$. 
\begin{table}[tb]
	\centering
	\begin{tabular}{|c|c|c||c|} \hline 
		\rule[14pt]{0pt}{0pt}	  &  $\tau\approx i$ in NH    &   $\tau\approx i\infty$ in NH &  IH\\ \hline 
		\rule[14pt]{0pt}{0pt}	
		$\tau$&   $ -10^{-4}+ 1.024 \, i$  & $0.0016+ 1.650 \, i$  & $0.00043+ 1.593\, i$ \\ 
		\rule[14pt]{0pt}{0pt}
		$[\frac{m_{H_1}}{\rm GeV},\frac{m_{H_2}}{\rm GeV},\frac{m_k}{\rm GeV}]$ 
		&$ [1826, 697.9, 226.4]$ & [705.8, 4198, 1282]  & [180.8, 744.7, 2030] \\
		\rule[14pt]{0pt}{0pt}
		$[\frac{m_{\eta_0}}{\rm GeV},\frac{M_0}{\rm GeV}]$ 
		&$ [493.5, 3470]$ & [ 958.3, 5983]  & [6557.72, 6563.72] \\
		\rule[14pt]{0pt}{0pt}
		$[\kappa_1,\kappa_2,\kappa_3]$ 
		& [0.122, -0.272, 0.212] & [0.0615, 0.252, -0.369]  & [0.0205, 0.0220, 0.0228] \\
		\rule[14pt]{0pt}{0pt} 
		$[\rho_1,\rho_2,\rho_3]$ 
		& $[-0.036, 0.44 e^{-4.5i}, 0.028e^{3.6i}]$ & $[0.07, 0.12e^{13i}, 0.14e^{8.2i}]$  & $[-0.776, 0.015 e^{-4.0i}, 0.030e^{0.77i}]$ \\
		\rule[14pt]{0pt}{0pt}
		$\sin^2\theta_{12}$ & $ 0.328$	& $ 0.345$ & $ 0.309$\\
		\rule[14pt]{0pt}{0pt}
		$\sin^2\theta_{23}$ &  $ 0.557$	& $ 0.575$ & $ 0.504$\\
		\rule[14pt]{0pt}{0pt}
		$\sin^2\theta_{13}$ &  $ 0.0239$	&  $ 0.0242$ &  $ 0.0227$\\
		\rule[14pt]{0pt}{0pt}
		$\delta_{CP}^\ell$ &  $251^\circ$ 	&  $252^\circ$ 	&  $183^\circ$\\
		\rule[14pt]{0pt}{0pt}
		$[\alpha_{21},\,\alpha_{31}]$ &  $[35.9^\circ,\,245^\circ]$ 	&$[197^\circ,\,269^\circ]$ 	&$[0.808^\circ,\,179^\circ]$\\	
		\rule[14pt]{0pt}{0pt}
		$\sum m_i$ &  $58.8$\,meV 	& $67.0$\,meV & $120$\,meV\\
		\rule[14pt]{0pt}{0pt}
		$\langle m_{ee} \rangle$ &  $3.50$\,meV 	&  $0.217$\,meV &  $50.1$\,meV \\
		\rule[14pt]{0pt}{0pt}
		$\langle m_\beta \rangle$ &  $9.21$\,meV 	&  $11.13$\,meV 	&  $51.9$\,meV \\
		\hline
	\end{tabular}
	\caption{Numerical values of parameters and observables
		at the sample points of NH and IH.}
	\label{samplelepton}
\end{table}


The other remarks are in order:
 \begin{enumerate}
\item 
We have found that allowed regions for three mixings run over the experimental ranges in Eq.~(\ref{eq:exp}) except the case of $\sin^2\theta_{23}$ in IH, therefore there are no correlations among observables. The allowed region of $\sin^2\theta_{23}$ in IH is [0.47--0.55], which is narrower than the experimentally allowed region.
\item 
We have found muon anomalous magnetic moment to be $1.2\times10^{-11}$ for NH and  $6.0\times10^{-15}$ for IH,
both of which are much below the experimentally favored region $\sim10^{-9}$. 
\item 
We have found upper limits for branching ratios of LFVs;  $4.0\times10^{-13}$ for $\mu\to e\gamma$, $3.0\times10^{-13}$ for $\tau\to e\gamma$, and $8.5\times10^{-13}$ for $\tau\to \mu\gamma$ in case of NH,
while $3.5\times10^{-17}$ for $\mu\to e\gamma$, $3.0\times10^{-19}$ for $\tau\to e\gamma$, and $4.0\times10^{-19}$ for $\tau\to \mu\gamma$ in case of IH. NH could be tested in the near future, while IH is far from the current bounds.
 \end{enumerate}

Here we briefly discuss possible collider signature of new particles in our model.
There are several charged particles in the model such that exotic charged lepton $E$, charged scalars $H^\pm_{1,2}$ and doubly charged scalar $k^{\pm \pm}$ 
that can be produced at collider experiments like the LHC~\footnote{New neutral scalar from $\varphi$ can be also produced via mixing with the SM Higgs. However we here omit to discuss signatures from this neutral scalar production since the production cross-section can be suppressed by considering small Higgs mixing.} . 
Assuming exotic lepton is heavier than singly charged scalar, it can decay as $E \to H^-_{1,2} \nu$  followed by decay of $H^-_{1,2} \to W^{- (*)} \eta_{R/I}$ where $\eta_{R/I}$ is DM candidate in our scenario. 
Also doubly charged scalar $k^{\pm \pm}$ can decay into $E^{\pm} E^{\pm}$ and/or $H_1^\pm H_1^\pm$. 
Then typical signatures from charged particle productions are found to be 
\begin{align}
& pp \to H_{1,2}^+ H_{1,2}^-  \to W^{+ (*)} W^{- (*)} \eta^0 \eta^0  \\
& pp \to E^- E^+ \to H_{1,2}^+ H_{1,2}^- \nu \bar \nu  \to W^{+ (*)} W^{- (*)} \nu \bar \nu \eta^0 \eta^0 \\ 
& pp \to k^{++} k^{--} \to  2H_{1,2}^+ 2H_{1,2}^-  \to 2 W^{+(*)} 2 W^{-(*)} 2 \eta^0 \\
& pp \to k^{++} k^{--} \to  2E^+ 2E^-  \to 2H_{1,2}^+ 2H_{1,2}^- 2\nu 2 \bar \nu \to 2 W^{+(*)} 2 W^{-(*)} 2 \nu 2 \bar \nu 4 \eta^0
\end{align}
where $\eta^0$ stands for $\eta_{R}$ or $\eta_I$.
Thus typical signature of our model will be multi $W$ bosons with missing transverse energy at the LHC. 
Detailed analysis of such signatures is beyond the scope of this paper and it is left as future work.

\section{Conclusion and discussion}
\label{sec:conclusion}

We have proposed a two-loop induced neutrino mass model with a modular $A_4$ symmetry, 
and discussed  predictions of neutrino oscillation data as well as LFVs, muon anomalous magnetic moment related to interactions generating neutrino mass. We have found several predictions about the lightest neutrino mass, neutrinoless double beta decay, sum of the neutrino masses, and phases for NH and IH, classifying each of the fixed point region nearby $\tau=i,i\times\infty$.
Here we would not found any solutions nearby $\tau=\omega$ that is also one of the fixed points.

Before closing this section, it would be worthwhile to mention dark matter  candidate.
In our model, neutral component of $\eta$ can be the one, if there is mass difference between the real part and imaginary part
to evade the direct detection search.
This can be achieved by introducing, e.g., $SU(2)_L$ triplet boson with $(1,-2)$ for hypercharge and modular weight, respectively.
The systematic analysis has already been done by ref.~\cite{Hambye:2009pw} that tells us the dark matter mass is at around 534 GeV when the mass is larger than the mass of $W/Z$ mass.
In lighter region, one also finds the dark matter mass is at around the half of Higgs mass; 63 GeV.

\section*{Acknowledgments}
\vspace{0.5cm}
{\it
This research is supported by the Ministry of Science, ICT and Future Planning, Gyeongsangbuk-do and Pohang City (H.O.). 
H. O. is sincerely grateful for the KIAS member, too.}


\end{document}